\documentclass[%
 preprint, amsmath,amssymb,aps, prf]{revtex4-1}
\usepackage{float}
\usepackage{graphicx}
\usepackage{dcolumn}
\usepackage{bm}
\usepackage{color}
\usepackage{amsmath}
\usepackage{amsthm}

\usepackage[colorlinks = true,
            linkcolor = blue,
            urlcolor  = blue,
            citecolor = blue,
            anchorcolor = blue]{hyperref}

\begin{document}
\title{Thermodynamic Coupling of Mass and Electromagnetic Fields: Entropic Origin of Parity Asymmetry and the Meissner Effect}
\author{Fei Wang}
\email{fei.wang@kit.edu}

\affiliation{
 Institute for Applied Materials - Microstructure Modelling and Simulation (IAM-MMS),
Karlsruhe Institute of Technology (KIT),
Strasse am Forum 7, 76131 Karlsruhe, Germany
}

\affiliation{Institute of Nanotechnology (INT),
Karlsruhe Institute of Technology (KIT),\\
Hermann-von-Helmholtz-Platz 1, 76344 Eggenstein-Leopoldshafen, Germany}

\date{\today}

% 
% 
% \author[inst1,inst2]{Fei Wang\fnref{label2}}
% % \cortext[correspondingauthor]{Corresponding author: fei.wang@kit.edu}
% \fntext[label2]{fei.wang@kit.edu}
% \affiliation[inst1]{organization={Institute of Applied Materials-Microstructure Modelling and Simulation, Karlsruhe Institute of Technology (KIT)},%Department and Organization
%             addressline={Strasse am Forum 7}, 
%             city={Karlsruhe},
%             postcode={76131}, 
%             country={Germany}}
% \affiliation[inst2]{organization={Institute of Nanotechnology, Karlsruhe Institute of Technology (KIT)},%Department and Organization
%             addressline={Hermann-von-Helmholtz-Platz 1}, 
%             city={Eggenstein-Leopoldshafen},
%             postcode={76344}, 
%             country={Germany}}

\begin{abstract}
We develop a thermodynamic framework that couples mass dynamics, described by the Newton-Gibbs-van der Waals formalism, with electromagnetic fields beyond the scope of classical Maxwell theory. 
Classical Newtonian mechanics does not capture density evolution in the momentum balance, 
while the standard Maxwell equations neglect the contribution of the curl component of the electric field associated with moving charges.
Building on an alternative understanding on entropy, 
we develop a generalized theory for electrodynamics governed by entropy-production constraints. 
The resulting framework yields a modified Maxwell stress tensor that incorporates the moving-charge contribution, leading to intrinsic parity asymmetry in electromagnetic forces.
The theory naturally reproduces key features of superconductivity, including the Meissner effect, and reduces to the conventional Maxwell-Faraday and Maxwell-Ampere equations in an appropriate limit. 
This entropic formulation provides a unified thermodynamic basis for mass-field coupling and reveals new physical consequences arising from motion-induced electromagnetic effects.
\end{abstract}

%\begin{graphicalabstract}

%Graphical abstract

%\end{graphicalabstract}

% \begin{graphicalabstract}
% \includegraphics[width=\textwidth]{27November_Graphical_abstract.jpg}
% \end{graphicalabstract}

% \begin{keyword}
%  Maxwell equations  \sep Electromagnetic Dynamics \sep Thermodynamic Theory \sep Entropy 
% \end{keyword}
% 
% \end{frontmatter}

% 
% \begin{document}
% 
% \title[Entropy concept for coupling theory of 
% electro-magnetic field with mass]
% {Thermodynamic Coupling of Mass and Electromagnetic Fields: Entropic Origin of Parity Asymmetry and the Meissner Effect}
% 
% \author{Fei Wang}
% \email{fei.wang@kit.edu}
% 
% \affiliation{Institute for Applied Materials,
% Karlsruhe Institute of Technology (KIT),
% Strasse am Forum 7, 76131 Karlsruhe, Germany
% }
% 
% \affiliation{Institute of Nanotechnology (INT),
% Karlsruhe Institute of Technology (KIT),\\
% Hermann-von-Helmholtz-Platz 1, 76344 Eggenstein-Leopoldshafen, Germany
% }
% 
% 
% \begin{abstract}

% \end{abstract}

 \maketitle

\section{Introduction}

In classical physics, mass systems and electromagnetic fields are treated within distinct frameworks. The behavior of mass systems is typically described using the Newton-Gibbs-van der Waals formulation~\cite{benacquista2018classical,rowlinson1979translation}, while the dynamics of electromagnetic fields are governed by the classical Maxwell equations~\cite{landau2013electrodynamics}. Both frameworks are well established and extensively studied, and it is not the aim of this work to provide a comprehensive review of their individual developments. Instead, we focus on the coupling between the Newton-Gibbs-van der Waals framework and Maxwell's equations. Establishing such a coupling remains an open problem, primarily due to the following challenges:

\begin{itemize}
 \item [(I):] Newtonian mechanics relates the force  $\boldsymbol{F}$
 to inertial through $m \boldsymbol{a}$ as $\boldsymbol{F}=m \boldsymbol {a}$, 
 where $m$ is the particle mass and $\boldsymbol{a}=\mathrm{d}\mathbf{u}/\mathrm{d}t$ 
 depicts the acceleration which is the total time derivative of the velocity $\mathbf{u}$. 
 However, within this framework, Newton's law does not account for the density evolution appearing in the momentum balance,
 $\mathrm{d} (\rho\mathbf{u})/\mathrm{d}t$.
 In my previous work~\cite{wang2024compressible}, this limitation was addressed by introducing a velocity entropy
 $s_u$ into  the Gibbs-van der Waals formulation,
 yielding an entropy interpretation of acceleration
 \begin{equation}
 \rho\mathbf{u}\cdot \frac{\mathrm{d}\mathbf{u} }{\mathrm{d} t}=-T\frac{\mathrm{d} s_u}{\mathrm{d} t}.
 \end{equation}
Here, $T$ is the absolute temperature and $\rho$ represents the density.
 \item [(II):] The Maxwell equations fail to capture the moving effect of charges.
 According to the Helmholtz decomposition, the electrical field $ \mathbf{E}$ can be decomposed
 into a curl-free scalar-field $\psi_0$ ($\nabla \times \nabla\psi_0=\bold 0$)
 and a divergence free vector-field $\boldsymbol\psi_1$ [$\nabla\cdot (\nabla\times \boldsymbol\psi_1)=0 $]:
\begin{equation}
 \mathbf{E}=\nabla \psi_0+\nabla\times \boldsymbol\psi_1.
\end{equation}
The classic Maxwell equations only consider the electrostatic part, $\nabla \psi_0$.
Specifically, the classic Maxwell stress tensor 
does not consider the moving effect associated with the vector field, $\boldsymbol\psi_1$.
\end{itemize}

In this work, I extend the entropy concept introduced in (I) for modifying Newtonian mechanics 
to the electromagnetic field in (II), and develop a generalized theory of electrodynamics. 
The proposed framework is built on a simple but fundamental principle of free-energy minimization or, equivalently, entropy maximization
\begin{equation}
 \frac{\mathrm{d}}{\mathrm{d}t}\int_\Omega  s \mathrm{d}\Omega\geq 0
 \end{equation}
by giving a new understanding on the entropy $s$.
Our theory offers several key advantages:
\begin{itemize}
 \item [(a)]
 It shows that the motion-induced contribution associated with 
$\boldsymbol\psi_1$  modifies the classical Maxwell stress tensor. This generalized stress tensor leads naturally to parity asymmetry in electromagnetic forces.
 \item [(b)] It is consistent with the Meissner effect in superconductors, 
 predicting zero electrical resistance when the magnetic field is zero.
\item  [(c)]  In an appropriate limiting case, it reproduces the classical Maxwell equations, including the
Maxwell-Faraday and Maxwell-Ampere laws~\cite{wang2022expanded}.
\item  [(d)]  It provides a general definition for the speed of  sound and light.
 \end{itemize}

The remainder of the paper is organized as follows. In Section II, we introduce a general dissipation principle. Section III presents the derivation of the time derivative of the energy functional for a mass system, while Section IV addresses the time derivative of the energy functional for electromagnetic fields. In Section V, we formulate the kinetic equations based on the dissipation principle. Section VI provides a case study demonstrating the application of the present theory, and Section VII concludes the work.

\section{Dissipation principle}
\label{sec:3}
\textit{Theorem:} We consider a domain $\Omega$ with boundary $\Gamma$.
The domain consists of a binary  mass system
and electromagnetic field.
We define a free energy functional as $\mathcal F(c): \Phi\mapsto\mathbb{R}$
and a composition function $c(t, \mathbf{x}(t)): \mathbb{R}^+\times\mathbb{R}^3\mapsto\mathbb{R}$.
We formulate the free energy functional as
\begin{equation}
 \mathcal F(c)=\int_\Omega f(c) \mathrm{d}\Omega=\int_\Omega (\theta-Ts_c)\mathrm{d}\Omega,
 \label{eq:1000}
\end{equation}
The physical meaning of the notations is as follows.
$\theta: \mathbb{R}^+\times\mathbb{R}^3\mapsto\mathbb{R}$ is the internal energy,
$T$ is the absolute temperature,
$s_c: \mathbb{R}^+\times\mathbb{R}^3\mapsto\mathbb{R} $ is the entropy associated with the composition $c$.

The total time derivative is expressed as 
\begin{equation}
\frac{ \mathrm{d} c(t, \mathbf{x}(t))}{\mathrm{d}t}=\frac{\partial c}{\partial t}+\mathbf{u}\cdot \nabla c,
\end{equation}
where $\mathbf{u}=\partial \mathbf{x}/\partial t$.

For the functional given by Eq.~\eqref{eq:1000}, 
we state the following dissipation principle to derive the kinetic equations. 
For a closed system, the following free energy dissipation law
\begin{equation}
\frac{ \mathrm{d}\mathcal F}{\mathrm{d}t}=\int_\Omega \frac{\mathrm{d} f}{\mathrm{d} c}\frac{\mathrm{d}c}{\mathrm{d}t}\mathrm{d}\Omega=
\int_\Omega \mu\frac{\mathrm{d}c}{\mathrm{d}t}\mathrm{d}\Omega\leq0
\label{eq:18-00}
\end{equation}
implies a time evolution equation for the conserved variable $c$:
\begin{equation}
 \frac{\mathrm{d}c}{\mathrm{d}t}=\nabla \cdot M_c\nabla \mu,
 \label{eq:19-00}
\end{equation}
and a time evolution equation for non-conserved variable $\hat{c}$:
\begin{equation}
 \frac{\mathrm{d}\hat{c}}{\mathrm{d}t}=-\tau  \mu, 
 \label{eq:20-00}
\end{equation}
where $M_c$ and $\tau$ are positive constants.

\begin{proof}[Proof of the dissipation principle:]
Substituting Eq.~\eqref{eq:19-00} 
into Eq.~\eqref{eq:18-00}, applying integration by parts,
and using the no-flux boundary condition, we obtain the free energy evolution with time as 
\begin{align}
\frac{ \mathrm{d}\mathcal F}{\mathrm{d}t}&=\int_\Omega \frac{\mathrm{d} f}{\mathrm{d} c}\bigg(\nabla \cdot M_c\nabla \frac{\mathrm{d} f}{\mathrm{d} c} \bigg)\mathrm{d}\Omega\\
&=\int_{\Gamma} \frac{\mathrm{d} f}{\mathrm{d} c}M_c\nabla \frac{\mathrm{d}f}{\mathrm{d} c} \cdot\mathbf{n} \mathrm{d}\Gamma
-\int_\Omega M_c \bigg(\nabla \frac{\mathrm{d} f}{\mathrm{d} c}\bigg)^2 \mathrm{d}\Omega\\
&=-\int_\Omega M_c (\nabla \mu)^2 \mathrm{d}\Omega\leq0.
\end{align}
Substituting Eq.~\eqref{eq:20-00} 
into Eq.~\eqref{eq:18-00}, we obtain 
\begin{equation}
\frac{ \mathrm{d}\mathcal F}{\mathrm{d}t}=-\int_\Omega \tau \mu^2 \mathrm{d}\Omega\leq0.
\end{equation}
\end{proof}

The system energy is conserved when the free energy 
is compensated by a component-associated entropy term $Ts_c$, namely, $e=f+Ts_c$
 and 
\begin{equation}
 \frac{\mathrm{d} }{\mathrm{d} t}\int_\Omega e  \mathrm{d}\Omega=0.
\end{equation}
This equality is equivalent to
\begin{equation}
 \mu \frac{\mathrm{d} c}{\mathrm{d} t}=-T\frac{\mathrm{d} s_c}{\mathrm{d} t}.
\end{equation}
The left hand side is responsible for 
the minimization of the free energy, which is equivalent to 
the maximization of the entropy on the right hand side.

The above dissipation principle can be further applied for vector fields $\mathbf{u}$ and $\boldsymbol\theta$,
where $\boldsymbol\theta$ is a function of $\mathbf{u}$, for example,  $\boldsymbol\theta=\rho \mathbf{u}$ denoting the momentum.
For a closed system, the following energy dissipation law
\begin{equation}
\frac{ \mathrm{d}\mathcal F}{ \mathrm{d}t}=\int_\Omega \boldsymbol\theta \cdot \frac{ \mathrm{d}\mathbf{u}}{ \mathrm{d}t} \mathrm{d}\Omega
\label{eq:25-00}
\end{equation}
implies a time evolution equation for the conserved variable $\mathbf{u}$
\begin{equation}
 \frac{\mathrm{d}\mathbf{u}}{\mathrm{d}t}=\nabla \cdot \tau_1 (\nabla \boldsymbol\theta+\nabla \boldsymbol\theta^T)+\nabla  \cdot \tau_2(\nabla\cdot \boldsymbol\theta)\mathbf I,
 \label{eq:26-00}
\end{equation}
and a time evolution equation for non-conserved variable $\hat{\mathbf{u}}$
\begin{equation}
 \frac{\mathrm{d}\hat{\mathbf{u}}}{\mathrm{d}t}=-\tau_0  \boldsymbol \theta, 
 \label{eq:27-00}
\end{equation}
where $\tau_0$,  $\tau_1$, and $\tau_2$ are positive materials dependent mobilities.

\begin{proof}[Proof of the energy dissipation for vectors:]
Substituting Eq.~\eqref{eq:26-00} 
into Eq.~\eqref{eq:25-00}, applying integration by parts,
and using the no-flux boundary condition, we obtain the energy dissipation as 
\begin{align}
\frac{ \mathrm{d}\mathcal F}{\mathrm{d}t}=-\int_\Omega \tau_1 \nabla \boldsymbol \theta: \nabla \boldsymbol \theta
+\tau_2 (\nabla\cdot \boldsymbol\theta)^2 \mathrm{d}\Omega\leq0.
\end{align}
Substituting Eq.~\eqref{eq:27-00} 
into Eq.~\eqref{eq:25-00}, we obtain 
\begin{equation}
\frac{ \mathrm{d}\mathcal F}{\mathrm{d}t}=-\int_\Omega \tau_0 \boldsymbol\theta^2 \mathrm{d}\Omega\leq0.
\end{equation}
\end{proof}

The same understanding for the minimization
of the free energy and the maximization of the entropy 
can be obtained when compensating
the free energy by an associated entropy, $T s_u$.
\begin{equation}
 \boldsymbol\theta \cdot \frac{ \mathrm{d}\mathbf{u}}{ \mathrm{d}t}=-T\frac{\mathrm{d} s_u}{\mathrm{d} t}.
\end{equation}

\section{Time derivative of the mass energy functional}
We formulate the free energy functional of the mass system as 
\begin{equation}
 E=E_f+E_p+E_k.
\end{equation}
The individual three contributions are listed as follows:
\begin{align}
 \text{Chemical free energy:}~ E_f&=\int_\Omega f(c,\nabla c) \mathrm{d}\Omega,\\
 \text{Pressure energy:}     ~ E_p&=\int_\Omega p_1 \mathrm{d}\Omega,\\
  \text{Kinetic energy:}     ~E_k&=\int_\Omega \rho \mathbf{u}\cdot \mathbf{u} \mathrm{d}\Omega.
\end{align}
Based on the van der Waals-Gibbs-Cahn theory~\cite{rowlinson1979translation,cahn1958free,cai2024chemo,ridl2018lattice}, we write the chemical free energy functional as 
\begin{equation}
 E_f=\int_\Omega f(c,\nabla c) \mathrm{d}\Omega=\int_\Omega f_0(c)+\frac{1}{2}\kappa(\nabla c)^2 \mathrm{d}\Omega,
 \end{equation}
 where $f_0$ is the bulk free energy and $\kappa$ represents the gradient energy coefficient.
 
By decomposing the entropy  $s$
into thermal entropy $s_T$, mixing entropy $s_c$, 
and velocity entropy $s_u$, 
we express the free energy density as 
\begin{equation}
 f=\theta_0+ \mu c-Ts_\phi-T s_u-Ts_T.
\end{equation}
The first part is the internal energy solely determined by the temperature.
The second part is the chemical contribution.
The last three terms depict the entropy.
 
Next, we calculate the time derivative of the system
energy functional to derive the kinetic equations.
The time derivative of the chemical free energy functional
reads
\begin{align}
 \frac{\mathrm{d}E_f}{\mathrm{d}t}=\int_\Omega \frac{\partial f}{\partial c}\frac{\mathrm{d} c}{\mathrm{d}t}+ \frac{\partial f}{\partial \nabla c}\cdot  \frac{\mathrm{d} \nabla c}{\mathrm{d}t} \mathrm{d}\Omega.
 \label{eq:59}
 \end{align}
The total time derivative of $\nabla c$ reads
\begin{equation}
 \frac{\mathrm{d}\nabla c}{\mathrm{d}t}=\frac{\partial \nabla c}{\partial t}+\mathbf{u}\cdot \nabla \nabla c=
  \nabla \frac{\partial c}{\partial t}+\mathbf{u}\cdot \nabla \nabla c,
\end{equation}
where $ \nabla \nabla c$ is the covariant derivative, similar to the Hessian matrix.
Thus, the second term in Equation~\eqref{eq:59}
is rewritten as
\begin{align}
\int_\Omega \frac{\partial {f}}{\partial \nabla c}\cdot  \frac{\mathrm{d} \nabla c}{\mathrm{d}t} \mathrm{d}\Omega
=\int_\Omega \frac{\partial {f}}{\partial \nabla c}\cdot \bigg(\nabla \frac{\partial c}{\partial t}\bigg) +
\frac{\partial {f}}{\partial \nabla c}\cdot (\mathbf{u}\cdot \nabla \nabla c)
\mathrm{d}\Omega.
\label{eq:5555}
\end{align}
By using integration by parts and using no-flux boundary condition, 
the first term in Equation~\eqref{eq:5555}
is rewritten as 
\begin{align}
\int_\Omega \frac{\partial {f}}{\partial \nabla c}\cdot \nabla \frac{\partial c}{\partial t} \mathrm{d}\Omega &=\int_\Omega - 
 \bigg(\nabla \cdot\frac{\partial {f}}{\partial \nabla c}\bigg)\bigg(\frac{\mathrm{d}c}{\mathrm{d}t}-\mathbf{u}\cdot \nabla c\bigg)\mathrm{d}\Omega.
\end{align}
The second term in Equation~\eqref{eq:5555}
is written as
\begin{align}
\int_\Omega 
\frac{\partial {f}}{\partial \nabla c}\cdot (\mathbf{u}\cdot \nabla \nabla c)
\mathrm{d}\Omega=
\int_\Omega 
\bigg[\bigg(\nabla \cdot\frac{\partial f}{\partial \nabla c}\bigg)\cdot \nabla  \bigg]\nabla c\cdot \mathbf{u}
  \mathrm{d}\Omega.
\label{eq:66666}
\end{align}
Summarizing the above derivations
and  using 
the following vector calculus:
\begin{equation*}
 \nabla \cdot (\boldsymbol a \otimes \boldsymbol b)= (\nabla \cdot \boldsymbol a)\boldsymbol b+(\boldsymbol a\cdot \nabla) \boldsymbol b,
\end{equation*}
with $\bold a=\frac{\partial {f}}{\partial \nabla c}=\kappa \nabla c$
and $\bold b=\nabla c$,
we obtain the final expression for the time derivative of 
the chemical free energy functional $E_f$ as 
 \begin{align}
 \label{eq:32-000}
  \frac{\mathrm{d}{E_f} }{\mathrm{d}t}&=\int_\Omega \bigg[\bigg(\frac{\partial {f}}{\partial c}-\nabla \cdot\frac{\partial {f}}{\partial \nabla c}  \bigg)\frac{\mathrm{d}c}{\mathrm{d}t}+ 
 \nabla \cdot \bigg( \frac{\partial {f}}{\partial \nabla c}\otimes \nabla c\bigg)\cdot \mathbf{u}\bigg]\mathrm{d}\Omega\\
 &=\int_\Omega \{\mu\frac{\mathrm{d}c}{\mathrm{d}t}+ 
 [\nabla \cdot ( \kappa\nabla c\otimes \nabla c)]\cdot \mathbf{u} \}\mathrm{d}\Omega,
 \end{align}
 where, according to Eq.~\eqref{eq:32-000},
 the chemical potential $\mu$ is defined as 
 \begin{equation}
  \mu= \frac{\partial {f}}{\partial c}-\nabla \cdot\frac{\partial {f}}{\partial \nabla c}.
  \label{eq:34-5}
 \end{equation}

 The time derivative of the mass pressure energy reads
\begin{equation}
 \frac{\mathrm{d} E_p}{\mathrm{d}t}=\int_\Omega \frac{\mathrm{d}p_1}{\mathrm{d}\mathbf r}\cdot \frac{\mathrm{d}\mathbf r }{\mathrm{d}t} \mathrm{d}\Omega
 =\int_\Omega \nabla p_1\cdot \mathbf{u} \mathrm{d}\Omega.
\end{equation}

The time derivative of the mass kinetic energy reads
\begin{equation}
 \frac{\mathrm{d} E_k}{\mathrm{d}t}=\int_\Omega \bigg[\frac{\mathrm{d} (\rho \mathbf{u})}{\mathrm{d}t}\cdot \mathbf{u} +(\rho \mathbf{u})\cdot \frac{\mathrm{d}  \mathbf{u}}{\mathrm{d}t} \bigg]\mathrm{d}\Omega.
\end{equation}
 
\section{Time derivative of the  energy functional
of electromagnetic fields}

The electromagnetic potential energy is 
integrated over the electromagnetic free energy density $f_b$ as 
  \begin{equation}
       \mathcal E_b=\int_\Omega  f_b\mathrm{d}\Omega
       =\int_\Omega  \frac{1}{2}(\mathbf{D}\cdot \mathbf E+ \mathbf B\cdot \mathbf H)\mathrm{d}\Omega,
  \end{equation} 
where $\mathbf E$
and $\mathbf H$ are the displacements of the electrical and magnetic fields, respectively;  $\mathbf{D}$
and $\mathbf{B}$ are the associated forces
following the constitutive relations: 
\begin{equation}
 \mathbf D=\varepsilon \mathbf{E},\ \mathbf B=\chi \mathbf{H}.
\end{equation}
where $\varepsilon$ is  electrical permittivity
and $\chi$ denotes magnetic permeability. 

An important modification here is the general formulation
for the displacements of the electrical field $\mathbf E$
and the magnetic field $\mathbf H$:
\begin{align}
 \mathbf E=\nabla\psi_0+\nabla\times \boldsymbol\psi_1;\
 \mathbf H=\nabla\varphi_0+\nabla\times \boldsymbol\varphi_1.
  \label{eq:4-0}
\end{align}
 Here,  $\psi_0$  and  $\varphi_0 $
      refer to the static potentials of
      electrical and magnetic fields.
Noteworthily, we introduce two new vectors,
      $\boldsymbol\psi_1$ and  $\boldsymbol\varphi_1$
      to 
      depict the moving effect of the electromagnetic fields that have not been  considered in the classic Maxwell equations.
      We will see 
      these new vector terms
      lead to broken symmetry of parity.
      When $\mathbf E$ and $\mathbf H$ are continuous functions in the space,
the validity of Equation~\eqref{eq:4-0}
  is ensured by the Helmholtz decomposition theory~\cite{sprossig2010helmholtz}.

The electromagnetic free energy $\mathcal E_b$ plus the kinetic energy $\mathcal E_k$ and pressure 
energy $\mathcal E_p$
is expressed as 
\begin{align}
 \mathcal E=  \mathcal E_b+ \mathcal E_p+  \mathcal E_k,
 \end{align}
with the following expressions for each part:
\begin{align}
 \mathcal E_b&=\mathcal E_{e_1}+\mathcal E_{e_2},\   
 \mathcal E_{e_1}=\int_\Omega\frac{1}{2} (\mathbf{D}\cdot \mathbf E)\mathrm{d}\Omega,\
  \mathcal E_{e_2}=\int_\Omega\frac{1}{2} (\mathbf B\cdot \mathbf H )\mathrm{d}\Omega,\\
  \mathcal E_k&= \int_\Omega (\mathbf{B}\times \mathbf{D}) 
 \cdot \mathbf{v}\mathrm{d}\Omega,\   \mathcal E_p=  \int_\Omega  p_2\mathrm{d}\Omega.
\end{align}
The formulation of the kinetic energy
is motivated by the Poynting vector.
Here,
we distinguish the movement velocity of the mass, $\mathbf{u}$,
from the propagation speed of the electromagnetic field, $\mathbf{v}$, 
unlike assuming $\mathbf{u}=\mathbf{v}$ in Ref.~\cite{zhang2024multi}.

Next, we derive the total time derivative
of the system energy as 
\begin{align}
 \frac{\mathrm{d}\mathcal E_{e_1}}{\mathrm{d}t}
=\int_\Omega  \frac{1}{2}\varepsilon^\prime \mathbf{E}^2 \frac{\mathrm{d}c}{\mathrm{d}t} 
 + \mathbf{D}\cdot \frac{\mathrm{d} \nabla \psi_0}{\mathrm{d} t}
 +\mathbf{D}\cdot \frac{\mathrm{d}\nabla \times \boldsymbol\psi_1}{\mathrm{d} t}.
\label{eq:153-000}
\end{align}
The first term in Equation~\eqref{eq:153-000}
can be adsorbed into the mass effect term [Eq.~\eqref{eq:34-5}],
leading to a modified chemical potential plus $\frac{1}{2}\varepsilon^\prime \mathbf{E}^2$.

Applying the divergence theory to the second term in Equation~\eqref{eq:153-000}
and no-flux boundary condition
leads to
\begin{align}
 \int_\Omega 
  \mathbf{D}\cdot \frac{\mathrm{d} \nabla \psi_0}{\mathrm{d} t}
  =\int_\Omega \{-(\nabla\cdot  \mathbf{D}) \frac{\mathrm{d} \psi_0}{\mathrm{d}t}
  + [\nabla \cdot ( \mathbf{D}\otimes \nabla\psi_0)]\cdot \mathbf{v}\}\mathrm{d} \Omega.
\end{align}
Applying the divergence theory to the third term in Equation~\eqref{eq:153-000}
and no-flux boundary condition
results in 
\begin{align}
 \int_\Omega 
  \mathbf{D}\cdot \frac{\mathrm{d} \nabla \times \boldsymbol\psi_1}{\mathrm{d} t}
  =\int_\Omega \{(\nabla\times  \mathbf{D}) \cdot \frac{\partial \boldsymbol\psi_1}{\partial t}
  + \nabla \cdot [ \mathbf{D}\otimes (\nabla \times \boldsymbol\psi_1)]\cdot
  \mathbf{v}
 -(\nabla \cdot \mathbf{D})(\nabla \times \boldsymbol\psi_1)\cdot \mathbf{v}\}
\mathrm{d}\Omega.
\end{align}
Therefore, we obtain the final expression for the time derivative of the electrical free energy as 
\begin{align}
 \frac{\mathrm{d} \mathcal E_{e_1}}{\mathrm{d}t}=
\int_\Omega \{-(\nabla\cdot  \mathbf{D}) \frac{\mathrm{d} \psi_0}{\mathrm{d}t}
+ (\nabla\times  \mathbf{D}) \cdot \frac{\partial \boldsymbol\psi_1}{\partial t}
+ [\nabla\cdot (\mathbf{D}\otimes\mathbf{E})]\cdot \mathbf{v} -(\nabla \cdot \mathbf{D})(\nabla \times \boldsymbol\psi_1)\cdot \mathbf{v} \}\mathrm{d}\Omega.
\label{eq:57-99}
\end{align}

Applying the same derivation to the magnetic part results in 
\begin{align}
 \frac{\mathrm{d}\mathcal E_{e_2}}{\mathrm{d}t}
=\int_\Omega  \frac{1}{2}\chi^\prime \mathbf{H}^2 \frac{\mathrm{d}c}{\mathrm{d}t} 
 + \mathbf{B}\cdot \frac{\mathrm{d} \nabla \varphi_0}{\mathrm{d} t}
 +\mathbf{B}\cdot \frac{\mathrm{d}\nabla \times \boldsymbol\varphi_1}{\mathrm{d} t}.
\label{eq:153-0}
\end{align}
which gives rise to a formulation quite similar to Eq.~\eqref{eq:57-99} as 
\begin{align}
 \frac{\mathrm{d}\mathcal E_{e_2}}{\mathrm{d}t}=
\int_\Omega \{-(\nabla\cdot  \mathbf{B}) \frac{\mathrm{d} \varphi_0}{\mathrm{d}t}
+ (\nabla\times  \mathbf{B}) \cdot \frac{\partial \boldsymbol\varphi_1}{\partial t}
+ [\nabla\cdot (\mathbf{B}\otimes\mathbf{H})]\cdot \mathbf{v} -(\nabla \cdot \mathbf{B})(\nabla \times \boldsymbol\varphi_1)\cdot \mathbf{v} \}\mathrm{d}\Omega.
\label{eq:49-99}
\end{align}
As $\nabla \cdot  \mathbf{B}=0$ (see Section VA), the moving effect of the magnetic field 
$\nabla \times \boldsymbol\varphi_1$ vanishes
and the last term in Eq.~\eqref{eq:49-99} is zero.
However, as  $\nabla \cdot  \mathbf{D}\neq0$ in media with charges, 
the last term in Eq.~\eqref{eq:57-99}
does not vanish; this term depicts the 
movement effect of the electrical field,
which is the key reason for parity asymmetry (see Section VI E).

The time derivative of the electromagnetic pressure energy reads
\begin{equation}
 \frac{\mathrm{d} \mathcal E_p}{\mathrm{d}t}=\int_\Omega \frac{\mathrm{d}p_2}{\mathrm{d}\mathbf r}\cdot \frac{\mathrm{d}\mathbf r }{\mathrm{d}t} \mathrm{d}\Omega
 =\int_\Omega \nabla p_2\cdot \mathbf{v} \mathrm{d}\Omega.
\end{equation}

The time derivative of the kinetic energy of the electric-magnetic field  reads
\begin{equation}
 \frac{\mathrm{d}\mathcal E_k}{\mathrm{d}t}=\int_\Omega \bigg[\frac{\mathrm{d} ( \mathbf{B}\times \mathbf{D})}{\mathrm{d}t}\cdot \mathbf{v} +(\mathbf{B}\times \mathbf{D})\cdot \frac{\mathrm{d}  \mathbf{v}}{\mathrm{d}t} \bigg]\mathrm{d}\Omega.
\end{equation}

\section{The kinetic equations}
For a coupled system
of mass and electromagnetic field, 
the total free energy is expressed as 
\begin{equation}
 \mathcal E=E_f+E_p+E_k+\mathcal E_b +\mathcal E_p+ \mathcal E_k. 
\end{equation}
For a closed system, the total free energy dissipates with time
\begin{equation}
\frac{\mathrm{d}\mathcal E}{\mathrm{d}t}\leq0.
\end{equation}
The total energy is conserved when an entropy contribution is included.
\begin{align}
 E =\mathcal E+T\int_\Omega (s_T+s_c+s_u+ s_{\psi_0}+s_{\psi_1}+s_{\varphi_0}+s_{\varphi_1}+s_v)\mathrm{d}\Omega,
\end{align}
which leads to different types of kinetic equations
\begin{align}
\label{eq:57-1}
 \mu\frac{\mathrm{d}c }{\mathrm{d}t}&=-T\frac{\mathrm{d} s_c}{\mathrm{d} t},\\
 \label{eq:57-2}
 \rho \mathbf{u}\cdot\frac{\mathrm{d} \mathbf{u} }{\mathrm{d}t}&=-T\frac{\mathrm{d} s_u}{\mathrm{d} t},\\
  \label{eq:57-3}
  (\mathbf{B}\times \mathbf{D})\cdot\frac{\mathrm{d} \mathbf{v} }{\mathrm{d}t}&=-T\frac{\mathrm{d} s_v}{\mathrm{d} t},\\
  \label{eq:57-4}
(\nabla \cdot \mathbf{D}) \frac{\mathrm{d} \psi_0 }{\mathrm{d}t}&=-T\frac{\mathrm{d} s_{\psi_0}}{\mathrm{d} t},\\
  \label{eq:57-5}
(\nabla \cdot \mathbf{B}) \frac{\mathrm{d} \varphi_0 }{\mathrm{d}t}&=-T\frac{\mathrm{d} s_{\varphi_0}}{\mathrm{d} t},\\
  \label{eq:57-6}
(\nabla \times \mathbf{D})\cdot \frac{\partial \boldsymbol\psi_1 }{\partial t}&=-T\frac{\partial s_{\psi_1}}{\partial t},\\
  \label{eq:57-7}
(\nabla \times \mathbf{B}) \cdot \frac{\partial \boldsymbol\varphi_1 }{\partial t}&=-T\frac{\partial s_{\varphi_1}}{\partial t}.
\end{align}

\subsection{Evolution equations}
According to the dissipation principle,
we obtain the  mass diffusion equations based on Eq.~\eqref{eq:57-1} as
\begin{equation}
 \frac{\mathrm{d} c}{\mathrm{d} t}=\nabla \cdot M_c \nabla \mu.
\end{equation}

According to Eq.~\eqref{eq:57-2}-Eq.~\eqref{eq:57-3},
the kinetic equations 
for the velocities $\mathbf{u}$ and $\mathbf{v}$ 
read
\begin{align}
 \frac{\mathrm{d} \mathbf{u}}{\mathrm{d} t}&= - \tau_u  (\rho\mathbf{u}),\\
 \frac{\mathrm{d} \mathbf{v}}{\mathrm{d} t}&=- \tau_v (\bold B\times  \bold D ).
\end{align}
or in the conserved way
\begin{align}
 \frac{\mathrm{d} \mathbf{u}}{\mathrm{d} t}&= \nabla \cdot \tau_u  \nabla (\rho\mathbf{u}),\\
 \frac{\mathrm{d} \mathbf{v}}{\mathrm{d} t}&= \nabla \cdot  \tau_v\nabla (\bold B\times  \bold D ).
\end{align}

Eq.~\eqref{eq:57-4}-Eq.~\eqref{eq:57-7} indicate the following electromagnetic equilibrium
\begin{align}
 \nabla\cdot \mathbf{B}&=0,\\
 \label{eq:68-8}
 \nabla\cdot \mathbf{D}&=-\rho_e,\\
  \nabla\times \mathbf{B}&=\bold 0,\\
 \nabla\times\mathbf{D}&=\bold 0.
\end{align}
Here, we have assumed that the time scale for the entropy relaxation 
of $s_{\psi_0}$, $s_{\varphi_0}$, $s_{\psi_1}$, and $s_{\varphi_1}$
is much smaller than that of other entropies.
The parameter $\rho_e$
depicts the charge density.

\subsection{Momentum conservation of mass and electromagnetic field}

Collecting all the $\cdot \mathbf{u}$ terms,
we obtain the momentum conservation for mass as 
\begin{align}
\frac{\mathrm{d} (\rho \mathbf{u})}{\mathrm{d} t}=-\nabla
  p_1-\nabla\cdot(\kappa \nabla c\otimes \nabla c),
\end{align}
where the pressure $p_1$
is expressed by integrating the Gibbs-Duhem relation as~\cite{chen1998lattice}
\begin{equation}
 -p_1=f-\mu c.
\end{equation}

Collecting all the $\cdot \mathbf{v}$ terms,
we obtain the momentum conservation for electromagnetic field as 
\begin{align}
  \frac{\mathrm{d} (\bold B\times  \bold D)}{\mathrm{d} t}=-\nabla \cdot \boldsymbol \Gamma+\rho_e  \bold (\nabla \times \boldsymbol \psi_1).
  \label{eq:14}
\end{align}
where the Maxwell stress tensor reads
\begin{align}
\boldsymbol \Gamma=-\bigg(\frac{1}{2}\bold D\cdot \bold E+\frac{1}{2}\bold B\cdot \bold H \bigg)\mathbf{I}+ \bold D\otimes\bold E + \bold B\otimes\bold H.
\end{align}

\section{Case study, validation, and application}
 
\subsection{Replication of the classic Maxwell equations} 
We decompose the Maxwell stress tensor $\boldsymbol \Gamma$ into two parts: 
 \begin{equation}
  \boldsymbol \Gamma=\boldsymbol {\mathcal{T}}_1+\boldsymbol {\mathcal{T}}_2,
 \end{equation}
with the electric part $\boldsymbol {\mathcal{T}}_1$ and magnetic part $\boldsymbol {\mathcal{T}}_2$ reading
\begin{align}
 \boldsymbol {\mathcal{T}}_1=-\frac{1}{2}\bold D\cdot \bold E
+  \bold D\otimes \bold E,\\
 \boldsymbol {\mathcal{T}}_2=-\frac{1}{2}\bold B\cdot \bold H
+  \bold B\otimes \bold H.
\end{align}
By using the vector calculus $\nabla\cdot (\bold a\otimes \bold b)$,
we calculate 
the divergence of the Maxwell-stress tensor $\boldsymbol {\mathcal{T}}_1$ as
\begin{align}
 \nabla \cdot  \boldsymbol {\mathcal{T}}_1
=-\frac{1}{2}  \bold E^2 \nabla\varepsilon  
   + (\nabla \cdot \bold D) \bold E
 -\bold D\times(\nabla \times \bold E).
  \label{eq:119}
\end{align}
The divergence of the Maxwell stress tensor $\boldsymbol {\mathcal{T}}_2$
reads
\begin{align}
 \nabla \cdot  \boldsymbol {\mathcal{T}}_2= -\frac{1}{2}(\nabla \chi)  \bold H^2
   + (\nabla \cdot  \bold B) \bold H
 - \bold B\times(\nabla \times \bold H)+ \bold J\times \bold B+\bold B\times \bold J,
    \label{eq:123}
\end{align}
where we have added a zero term, $\bold J\times \bold B+\bold B\times \bold J=\bold 0$;
$\bold J$ is the electrical current.

The total time derivative of the electromagnetic momentum reads
\begin{equation}
 \frac{\mathrm{d} (\bold B\times\bold D)}{\mathrm{d}t}=\bold B\times \frac{\mathrm{d} \bold D}{\mathrm{d}t}
 -\bold D\times \frac{\mathrm{d} \bold B}{\mathrm{d}t}.
 \label{eq:83-9}
\end{equation}

Collecting all the $\bold B\times$ terms in Equation~\eqref{eq:83-9} and Equation~\eqref{eq:123},
we obtain  
the  Maxwell-Ampere equation  as 
\begin{equation}
 \frac{\mathrm{d} \bold D}{\mathrm{d}t}=\nabla \times \bold H-\mathbf J.
 \label{eq:80-9}
\end{equation}
Collecting all the $\bold D\times$ terms in Equation~\eqref{eq:83-9} and Equation~\eqref{eq:119}, we obtain 
 the Maxwell-Faraday equation as 
\begin{equation}
 \frac{\mathrm{d} \bold B}{\mathrm{d}t}=-\nabla \times\bold E.
 \label{eq:81-9}
\end{equation}

The rest terms in Equation~\eqref{eq:123} and Equation~\eqref{eq:119}
are
\begin{align}
\notag
 \boldsymbol \Pi&=[\nabla \cdot (\varepsilon \bold E)] \bold E+ \bold J\times \bold B -\frac{1}{2}  \bold E^2\nabla \varepsilon-\frac{1}{2}  \bold H^2\nabla \chi\\
 &=\rho_e \bold E+ \bold J\times \bold B-\frac{1}{2}  \bold E^2\nabla \varepsilon  -\frac{1}{2}  \bold H^2\nabla \chi=\bold 0.
 \label{eq:82-9}
\end{align}
This vector term  $\boldsymbol \Pi$ must be a zero vector 
 to ensure the momentum conservation of electromagnetic field.
 
\subsection{Gauss's law}
Performing a divergence operator to both sides of Eq.~\eqref{eq:80-9},
we obtain the following steady state equation
\begin{equation}
 \mathbf{v}\cdot \nabla (\nabla \cdot \mathbf{D})+ \mathbf{u}\cdot \nabla \rho_e=0,
\end{equation}
where we have used the equality $\nabla\cdot \mathbf{u}=0$.
When $\mathbf{v}=\mathbf{u}$,
we obtain the Gauss's law, $\nabla \cdot \mathbf{D}=-\rho_e$ for the electrical field,
which is consisent with Eq.~\eqref{eq:68-8}.
Similarly, performing a divergence operator to both sides of Eq.~\eqref{eq:81-9},
we derive the Gauss's law, $\nabla \cdot \mathbf{B}=0$ for the magnetic field.

\subsection{Meissner effect}
 Recalling the above derivation and neglecting the term $\rho_e(\nabla\times \boldsymbol\psi_1)$
 in Eq.~\eqref{eq:14},
we see that 
the electromagnetic momentum evolution 
$\mathrm{d}( \bold B\times \bold D)/\mathrm{d}t=-\nabla\cdot\boldsymbol { \Gamma}$ 
leads to two subsequent effects:
\begin{itemize}
 \item [(I)]  Maxwell-Faraday and Maxwell-Ampere equations, which 
are responsible for the exchange of $\bold B$  and $\bold E$;
\item [(II)] A force term $ \boldsymbol \Pi$ given by  Eq.~\eqref{eq:82-9}.
\end{itemize}
To ensure the momentum conservation of electromagnetic fields
and to obtain the Maxwell-Faraday and Maxwell-Ampere equations,
 the following equality
\begin{equation}
\rho_e \bold E+ \bold J\times \bold B-\frac{1}{2}  \bold E^2\nabla \varepsilon  -\frac{1}{2}  \bold H^2\nabla \chi=\bold 0
 \label{eq:47-00}
\end{equation}
must hold everywhere in the domain.
In vacuum, there is no charge and no electric current, i.e., $\rho_e=0$
and $\mathbf{J}=\bold  0$. Therefore, Eq.~\eqref{eq:47-00} is automatically fulfilled in vacuum.
In media at the superconducting state, there is almost no dissipation
of the velocities $\mathbf{u}$
and $\mathbf{v}$.
The balance for the electron movement in the direction parallel to $\bold E$
and perpendicular to $\bold E$
reads
\begin{align}
 \rho_e E&=\frac{1}{2} \frac{\mathrm{d}\varepsilon}{\mathrm{d}x}E^2;\\
 JB&=\frac{1}{2} \frac{\mathrm{d}(1/\chi)}{\mathrm{d}x}B^2.
\end{align}
Combing these two equations, we obtain the following condition
for the movement of electrons
\begin{equation}
\label{eq:174}
\frac{J}{\rho_e}=\frac{\mathrm{d}(1/\chi)}{\mathrm{d}\varepsilon}\frac{B}{E}\Leftrightarrow  \frac{J}{\rho_e}=\frac{C^2 }{E}B,
\end{equation}
where $C=\sqrt{1/(\varepsilon\chi)}$ denotes the speed of light in media.
Due to axi-symmetry, the vectors, $\bold E$, $\bold B$, and $\bold J$ 
are represented by the scalar fields $E$, $ B$, and $ J$, respectively. 

From Eq.~\eqref{eq:174},
 the electrical resistance $r$ at any position is expressed as
 \begin{equation}
  r=\frac{E}{J}=\frac{\rho_e C^2 }{J^2}B.
 \end{equation}
When $B=0$, there is no resistance and we have superconducting.
The superconducting with  $B=0$ is consisent with the Meissner effect~\cite{meissner1933neuer,gavish2021ginzburg}.

\subsection{Speed of light}
\label{sec:6}
In vacuum, we have $\rho_e=0$ and therefore $\mathbf{J}=0$.
 By considering these two conditions, the Maxwell-Ampere equation
 is modified as 
 \begin{equation}
   \chi_0 \varepsilon_0 \frac{\mathrm{d} \mathbf{E} }{\mathrm{d}t}=\nabla\times   \mathbf{B},
   \label{eq:53}
 \end{equation}
where $ \chi_0$ and $\varepsilon_0$
are the magnetic permeability and electrical permittivity, respectively,
in vacuum.
By taking $\nabla\times$ on both sides of Eq.~\eqref{eq:53}, we obtain
\begin{equation}
 \chi_0 \varepsilon_0\frac{\mathrm{d}^2\mathbf{ B}}{\mathrm{d}t^2}=\nabla^2\mathbf{ B}.
 \label{eq:54}
\end{equation}

By using the total derivative for a one-dimensional setup without loss of generality
in $x$-coordinate,
\begin{align}
 \frac{\mathrm{d}B}{\mathrm{d}t}=\frac{\partial B}{\partial t}+ \frac{\mathrm{d}B}{\mathrm{d}x}\frac{\mathrm{d}x}{\mathrm{d}t}=\frac{\partial B}{\partial t}+ v\frac{\mathrm{d}B}{\mathrm{d}x},
 \end{align}
 we obtain the second derivative of the magnetic field as 
  \begin{align}
 \frac{\mathrm{d}^2B}{\mathrm{d}t^2}=\frac{\partial^2 B}{\partial t^2}+\frac{\mathrm{d} v}{\mathrm{d}t}\frac{\mathrm{d}B}{\mathrm{d}x}+2v\frac{\partial}{\partial t}\frac{\mathrm{d}B}{\mathrm{d}x}+v^2\frac{\mathrm{d}^2B}{\mathrm{d}x^2}.
\end{align}
For the steady state, $\partial X/\partial t=0$, $X=B,\ \mathrm{d}B/\mathrm{d}x$
and assuming that there is no acceleration, $\mathrm{d}v/\mathrm{d}t=0$, we obtain 
$\mathrm{d}^2B/\mathrm{d}t^2=v^2\mathrm{d}^2B/\mathrm{d}x^2$
and hence the speed of light in vacuum as 
\begin{equation}
 v^2=C^2=\frac{1}{\chi_0\epsilon_0},\quad \text{or} \quad C=\sqrt{\frac{1}{\chi_0\epsilon_0}}.
\end{equation}
Similarly, we can also derive the speed of light in media.

We propose a more general formulation to define the speed of light.
At the steady state, the momentum balance equation reduces to 
\begin{equation}
 \mathbf{v}\cdot \nabla (\mathbf{B}\times \mathbf{D})=-\nabla p_2
\end{equation}
Similar to the treatment of mass momentum, we assume the electromagnetic momentum takes the form
\begin{equation}
 \mathbf{B}\times \mathbf{D}=\rho_p \mathbf{v},
\end{equation}
where $\rho_p$ represents an effective ``mass density'' of the electromagnetic field, for instance, associated with photons.
Under this assumption, we obtain a general expression for the speed of light that is directly analogous to the formulation of the speed of sound:
\begin{equation}
 v=\sqrt{\frac{\mathrm{d} p_2}{\mathrm{d} \rho_p} }.
\end{equation}

\subsection{Sound speed}
Neglecting the Korteweg stress tensor in the momentum conservation of mass leads 
to 
\begin{equation}
 \rho\frac{\mathrm{d}\mathbf{u}}{\mathrm{d} t}+ \mathbf{u}\frac{\mathrm{d}\rho}{\mathrm{d} t}=-\nabla p_1.
 \label{eq:95}
\end{equation}
At isentropic condition implying that 
the time scale of $\mathrm{d} s_u/\mathrm{d} t$ is much less than $\mathrm{d} \rho/\mathrm{d} t$,
we have 
\begin{equation}
\frac{\mathrm{d}\mathbf{u}}{\mathrm{d} t}=0.
\label{eq:96}
\end{equation}
Substituting Eq.~\eqref{eq:96}
into Eq.~\eqref{eq:95} and considering the steady state in 1D, we obtain 
\begin{equation}
 u^2\mathrm{d}\rho=-\mathrm{d}p_1.
\end{equation}
With the expression for the pressure $-p_1=f-\mu c\triangleq P$,
we obtain the formulation for the speed of sound as 
\begin{equation}
 u=\sqrt{\frac{\mathrm{d} P}{\mathrm{d} \rho}}.
\end{equation}

 \subsection{Parity asymmetry}
 
Parity (mirror) symmetry refers to invariance under spatial inversion, 
$\bold x\rightarrow -\bold x$~\cite{landau2013electrodynamics}.
A system preserves parity if its equations of motion remain unchanged under this reflection operation. 
If the physical force or field reverses sign under the parity transformation, 
the system exhibits parity violation or broken mirror symmetry.
In the classic concept~\cite{landau1975classical}, 
$\bold B$ and $\bold H$ are even parity
with $\bold B(-\bold x)=\bold B(\bold x)$;
$\bold D$ and $\bold E$ are odd parity
with $\bold D(-\bold x)=-\bold D(\bold x)$.
This concept is mainly based on $\bold  E=\nabla \psi_0$
and $\bold  H=\nabla \times \boldsymbol\varphi_1$.
 As $\nabla_{-x}=-\nabla_x $, we have $\bold E(-\bold x)=-\bold E(\bold x)$.
As $\boldsymbol\varphi_1 (-\bold x)= -\boldsymbol\varphi_1 (\bold x)$,
we have $\bold H(-\bold x)=\bold H(\bold x)$.

From the momentum evolution of the electromagnetic field,
 we obtain a generalized electromagnetic force
 \begin{align}
 \label{eq:99-0}
&\boldsymbol f=-\nabla\cdot \boldsymbol \Gamma + \rho_e (\nabla \times \boldsymbol \psi_1);\\
& \boldsymbol\Gamma=-\big(\frac{1}{2}\bold D\cdot \bold E+\frac{1}{2}\bold B\cdot \bold H \big)\mathbf{I}+ \bold D\otimes\bold E + \bold B\otimes\bold H .
\end{align}
The first force term in Eq.~\eqref{eq:99-0}
is contributed by the classic Maxwell stress tensor.
This classic term gives rise to the symmetry of the parity.
Under a mirror (parity) transformation, 
$\bold D$ 
 and 
$\bold E$ both
change sign, becoming 
$-\bold D$ and  $-\bold E$.
However, the sign of  
$\bold D\otimes\bold E $ and $\bold D\cdot \bold E $  remains invariant under this transformation.
The momentum balance equation after applying the mirror operation becomes
\begin{equation}
 -\frac{\mathrm{d} (\mathbf{B}\times \mathbf D)}{\mathrm{d} t}=\nabla\cdot \boldsymbol \Gamma.
 \label{eq:103}
\end{equation}
The negative sign on the left-hand side arises from the sign reversal of $\mathbf{D}$ under the mirror transformation.
The electromagnetic force changes from 
$-\nabla\cdot \boldsymbol \Gamma$ to $\nabla\cdot \boldsymbol \Gamma$
is due to $\nabla_{-x}=-\nabla_x $.
As demonstrated in Eq.~\eqref{eq:103},
the mirror operation does not change the momentum balance equation.

The second term  in Eq.~\eqref{eq:99-0} leads to the breaking of parity symmetry.
In the mirror transformation,
the sign of $\nabla \times \boldsymbol \psi_1$
is unchanged, as $\nabla_{-x}=-\nabla_x $ leads to a negative sign
and $\boldsymbol \psi_1$ becomes $-\boldsymbol \psi_1$ (the vector $\boldsymbol \psi_1$ is an odd parity).
This results in the non-conservation of parity.
In fact, when $\bold E=\nabla \psi_0+ \nabla\times\boldsymbol \psi_1$,
the Maxwell stress tensor is not mirror symmetric anymore due to the cross term, $(\nabla\psi_0)\otimes (\nabla \times\boldsymbol\psi_1)$, which also leads 
to non-conservation of parity.

\section{Conclusion}

In summary, we have developed a generalized theory for the electromagnetic field based on 
the simple yet fundamental principle of entropy maximization, 
offering an alternative interpretation of entropy within the system. The proposed framework has been validated by its ability to reproduce the classical Maxwell equations, the propagation speeds of light and sound, and the Meissner effect in superconductors. A key outcome of the theory is a generalized electromagnetic force that includes not only the classical Maxwell stress tensor but also an additional term,
$\rho_e(\nabla \times \boldsymbol\psi_1)$, arising from the motion of charges. This new contribution provides a natural explanation for the non-conservation of parity in electromagnetic interactions.

\section*{Declaration of competing interest}

The author declares that he has no known competing financial interests or personal relationships that could have appeared to influence the work reported in this paper.

\section*{Data availability}

No data was used for the research described in the article.

\end{document}